\documentstyle[12pt]{article}

\newcommand{\newc}{\newcommand}
\newc{\beq}{\begin{equation}}
\newc{\eeq}{\end{equation}}
\newc{\bea}{\begin{array}}
\newc{\eea}{\end{array}}
\newc{\pd}{\partial}
\newc{\Psibar}{\overline\Psi}
\newc{\qbar}{\overline q}
\newc{\w}{{\bf w}}
\newc{\bfn}{{\mathbf\nabla}}
\newc{\bfg}{{\mathbf\gamma}}
\newc{\E}{{\mathbf{E}}}
\newc{\bp}{{\bf p}}
\newc{\la}{{\cal L}}
\newc{\ti}{{\times}}
\newc{\bA}{{\bf A}}
\newc{\ri}{{\rm i}}
\newc{\bL}{{\bf L}}
\newc{\bS}{{\bf S}}
\newc{\bB}{{\mathbf B}}
\newc{\bfx}{{\bf x}}
\newc{\bfV}{{\bf V}}
\newc{\bu}{{\bf u}}
\newc{\bM}{{\bf M}}
\newc{\bV}{{\bf V}}
\newc{\bv}{{\bf v}}
\newc{\bx}{{\bf x}}
\newc{\bD}{{\bf D}}
\newc{\bH}{{\bf H}}
\newc{\br}{{\bf r}}
\newc{\ve}{{\varepsilon}}
\newc{\bb}{{\bf b}}
\newc{\bc}{{\bf c}}
\newc{\bj}{{\bf j}}
\newc{\bd}{{\bf d}}
\newc{\bE}{{\mathbf{E}}}
\newc{\tla}{{\tilde{\cal L}}}
\newc{\ho}{\hookrightarrow }
\newc{\bP}{{\bf P}}
\newc{\piv}{{\partial_4}}
\newc{\pv}{{\partial_5}}
\newc{\bze}{{\mathbf 0}}

\newc{\sig}{{\mathbf\sigma}}
\newc{\bpi}{{\mathbf\pi}}
\newc{\eg}{{\rm e.g.\ }}
\newc{\ie}{{\rm i.e.\ }}
\newc{\etal}{{\rm et al\ }}
\def\JPA#1#2#3#4{#2 #1 {\it J. Phys. A: Math. Gen.} {\bf #3} #4}
\def\BFPS#1#2#3#4{#2 #1 {\it Brit. F. Phil. Sci.} {\bf #3} #4}
\def\PS#1#2#3#4{#2 #1 {\it Phil. Sci.} {\bf #3} #4}

\def\NGWG#1#2#3#4{#2 #1 {\it Nachr. Ges. Wiss. Gottingen} {\bf #3} #4}
\def\FP#1#2#3#4{#2 #1 {\it Fortschr. Phys.} {\bf #3} #4}
\def\MAG#1#2#3#4{#2 #1 {\it Math. Annalen} {\bf #3} #4}
\def\AP#1#2#3#4{#2 #1 {\it Ann. Phys. (NY)} {\bf #3} #4}
\def\CMP#1#2#3#4{#2 #1 {\it Comm. Math. Phys.} {\bf #3} #4}
\def\PLA#1#2#3#4{#2 #1 {\it Phys. Lett. A} {\bf #3} #4}
\def\APG#1#2#3#4{#2 #1 {\it Annalen Phys.} {\bf #3} #4}
\def\AHES#1#2#3#4{#2 #1 {\it Archive for History of Exact Sciences}
 {\bf #3} #4}

\def\SHPMP#1#2#3#4{#2 #1 {\it Studies in the History and Philosophy of Modern
 Physics} {\bf #3} #4}
\def\EPL#1#2#3#4{#2 #1 {\it Europhys. Lett.} {\bf #3} #4}
\def\AFLB#1#2#3#4{#2 #1 {\it Ann. Fondation Louis de Broglie} {\bf #3} #4}
\def\APP#1#2#3#4{#2 #1 {\it Acta Phys. Pol.} {\bf #3} #4}
\def\NC#1#2#3#4{#2 #1 {\it Nuov. Cim.} {\bf #3} #4}
\def\EPJB#1#2#3#4{#2 #1 {\it Eur. Phys. J.} {\bf B #3} #4}
\def\RPJ#1#2#3#4{#2 #1 {\it Russian Phys. J.} {\bf #3} #4}
\def\EJP#1#2#3#4{#2 #1 {\it Eur. J. Phys.} {\bf #3} #4}
\def\PTRSA#1#2#3#4{#2 #1 {\it Phil. Trans. R. Soc.} {\bf A #3} #4}
\def\JE#1#2#3#4{#2 #1 {\it J. Electrostatics} {\bf #3} #4}
\def\NCB#1#2#3#4{#2 #1 {\it Nuov. Cim.} {\bf B #3} #4}
\def\RSI#1#2#3#4{#2 #1 {\it Rev. Sc. Instrum.} {\bf #3} #4}
\def\CENTAURUS#1#2#3#4{#2 #1 {\it Centaurus} {\bf #3} #4}
\def\AIHP#1#2#3#4{#2 #1 {\it Ann. Inst. H. Poincar\'e} {\bf #3 B} #4}
\def\PRD#1#2#3#4{#2 #1 {\it Phys. Rev.} {\bf D #3} #4}
\def\PTRSL#1#2#3#4{#2 #1 {\it Phil. Trans. Roy. Soc. London} {\bf #3} #4}
\def\PR#1#2#3#4{#2 #1 {\it Phys. Rev.} {\bf #3} #4}
\def\IJTP#1#2#3#4{#2 #1 {\it Int. J. Theor. Phys.} {\bf #3} #4}
\def\JMP#1#2#3#4{#2 #1 {\it J. Math. Phys.} {\bf #3} #4}
\def\AJP#1#2#3#4{#2 #1 {\it Am. J. Phys.} {\bf #3} #4}
\def\RMP#1#2#3#4{#2 #1 {\it Rev. Mod. Phys.} {\bf #3} #4}
\catcode`@=11
\long
\def\@caption#1[#2]#3{\par\addcontentsline{\csname
  ext@#1\endcsname}{#1}{\protect\numberline{\csname
  the#1\endcsname}{\ignorespaces #2}}\begingroup
    \small
    \@parboxrestore
    \@makecaption{\csname fnum@#1\endcsname}{\ignorespaces #3}\par
  \endgroup}
\catcode`@=12

\begin{document}

\begin{titlepage}
\vskip 2cm
\begin{center}
{\Large\bf On the electrodynamics of moving bodies\\ at low velocities
\footnote{E-mail: {\tt montigny@phys.ualberta.ca,} 
 {\tt Germain.Rousseaux@inln.cnrs.fr}\hfill}}
\vskip 3cm
{\bf 
M. de Montigny$^{a,b}$ and G. Rousseaux$^{c}$ \\}
\vskip 5pt
{\sl $^a$Campus Saint-Jean, University of Alberta \\
 8406 - 91 Street \\
 Edmonton, Alberta, Canada T6C 4G9\\}
\vskip 2pt
{\sl $^b$Theoretical Physics Institute, University of Alberta\\
 Edmonton, Alberta, Canada T6G 2J1\\}
\vskip 2pt
{\sl $^c$Universit\'e de Nice Sophia-Antipolis, Institut Non-Lin\'eaire de Nice,\\
 INLN-UMR 6618 CNRS-UNICE, 1361 route des Lucioles,\\
 06560 Valbonne, France\\}
\vskip 2pt

\end{center}
\vskip .5cm
\rm
\begin{abstract}
We discuss the seminal article in which
 Le Bellac and L\'evy-Leblond have identified two Galilean
 limits of electromagnetism \cite{bellac}, and its modern
 implications.
 We use their results to point out some confusion in
 the literature and in the teaching of special
 relativity and electromagnetism. For instance, it
 is not widely recognized that there exist {\em two} well defined
 non-relativistic limits, so that researchers and teachers
 are likely to utilize an incoherent mixture of both.
 Recent works have shed a new light on the choice
 of gauge conditions in classical electromagnetism. We retrieve
 Le Bellac-L\'evy-Leblond's results by examining  
 orders of magnitudes, and then with a Lorentz-like
 manifestly covariant approach to Galilean covariance based
 on a 5-dimensional Minkowski manifold.
 We emphasize the Riemann-Lorenz 
 approach based on the vector and scalar potentials
 as opposed to the
 Heaviside-Hertz formulation in terms of electromagnetic fields.
 We discuss various applications and
 experiments, such as in magnetohydrodynamics and
 electrohydrodynamics, quantum mechanics, superconductivity, 
 continuous media, etc. Much of the current technology where
 waves are not taken into account, is actually based
 on Galilean electromagnetism. 
\end{abstract}
\end{titlepage}

\setcounter{footnote}{0} \setcounter{page}{1} \setcounter{section}{0} %
\setcounter{subsection}{0} \setcounter{subsubsection}{0}

\section{Introduction}

The purpose of this article is to emphasize the relevance of Galilean
 covariance in physics, even nowadays, about one
 hundred years after Lorentz, Poincar\'e
 and Einstein, then facing the apparent incompatibility
 between Galilean mechanics and the full set of Maxwell
 equations, have developed a theory that turned into special relativity \cite{darrigol1}.
 Seventy years later, Le Bellac and L\'evy-Leblond (LBLL) observed that there
 exist not only one, but {\em two} well-defined Galilean (that is,  
 non-relativistic) limits of electromagnetism: the so-called
 `magnetic' and `electric' limits \cite{bellac}. Although special relativity
 has superseded Galilean relativity when it comes to the description of
 high energy phenomena, there exists a wealth of low-energy systems,
 particularly in condensed matter physics and low-energy nuclear physics,
 where Galilean covariance should not be ignored. 

We wish to point out hereafter some confusion which results from not 
 recognizing appropriately the two Galilean limits of electromagnetism.
 This follows from inacurate definitions of non-relativistic covariance,
 which is why we emphasize at once
 that the definition of Galilean covariance
 employed henceforth in this paper rests on its compatibility with
 the Galilean transformations of space-time (Eq. (\ref{galxttrans}), below).   
 Examples of misleading, though well known, such text presentations are
 mentioned in \cite{bellac}, and there were many more since then.
 The fact that one should be careful when dealing with electrodynamics
 at low velocities has been illustrated, for instance,
 in Ref. \cite{crawford}. Let us illustrate this point with a simple
 example. Under a Lorentz transformation with relative velocity $\bv$,
 the electric and magnetic fields, in vacuum, become 
\beq\bea{l}
\bE'=\gamma(\bE+\bv\times\bB)+(1-\gamma)\frac{\bv (\bv\cdot\bE)}{\bv^2},\\
\bB'=\gamma(\bB-\frac 1{c^2}\bv\times\bE)+(1-\gamma)\frac{\bv (\bv\cdot\bB)}{\bv^2},
\eea\label{fieldslorentz}\eeq
respectively. 
The fact that Galilean covariance is a much more subtle concept than simply taking the
 $v<<c$, or $\gamma \simeq 1$, limit is illustrated by the fact that Eq. (\ref{fieldslorentz})
 then becomes:
\beq\bea{l}
\bE'=\bE+\bv\times\bB,\\
\bB'=\bB-\frac 1{c^2}\bv\times\bE,
\eea\label{wrongfieldtrans}\eeq
which not only is not compatible with Galilean relativity but, worse, does
 not even satisfy the composition properties of transformation groups
 \cite{bellac,crawford}. That is to
 say, a sequence of such transformations does not have the same form as above. 

We have organized this article as follows. In Section 2, we recall
 the main results of LBLL \cite{bellac} for later reference.
 In Section 3, we obtain these results using two arguments:
 one based on orders of magnitudes and a recent covariant approach
 with which the Galilean space-time is embedded into a five-dimensional
 space. Throughout the paper, we favour the
 Riemann-Lorenz formulation of electrodynamics, based on the scalar and vector
 potentials, over the Heaviside-Hertz approach which involved
 electromagnetic fields \cite{riemannlorenz}. Discussion and applications are in
 Section 4. Therein we present magnetohydrodynamics and
 electrohydrodynamics as two not-so-distant cousins which 
 can be associated respectively with the `magnetic' and
 `electric' Galilean limits of
 electrodynamics. We show that engineers are used to employ these Galilean limits
 which they denote as the electro- and magnetoquasistatics. We revisited Feynman's
 proof of the magnetic limit and illustrate the latter within the realm
 of superconductivity. Finally, we reassess our current understanding
 of the electrodynamics of moving bodies by examining the
 Trouton-Noble experiment in a Galilean context...
 
\section{Galilean electromagnetism}

 The purpose of LBLL was to write down the laws of
 electromagnetism in a form compatible
 with Galilean covariance rather than 
 Lorentz covariance. As LBLL put it, such laws could have been
 devised by a physicist in the mid-nineteenth
 century \cite{bellac}. Here, let us retrieve these laws from 
 relativistic kinematics. The Lorentz transformation of a four-vector
 $(u^0,\bu)$, where the four components have the same units,
 is given by (see chapter 7 of Ref. \cite{goldstein}):
\beq\bea{l}
u^{\prime 0}=\gamma\left (u^0-\frac 1c\bv\cdot\bu\right ),\\
\bu^\prime=\bu-\gamma\frac\bv cu^0+ (\gamma-1) \frac\bv{\bv^2}\bv\cdot\bu ,\eea
\label{lorentz4vector}\eeq
where $\gamma\equiv\frac 1{\sqrt{1-\bv^2/c^2}}$, 
with a relative velocity $\bv$. 
 The speed of light in the vacuum is denoted $c$.
 LBLL were the first to observe that this transformation
 admits two well-defined Galilean limits \cite{bellac}.
 One limit is for timelike vectors:
\beq\bea{l}
u^{\prime 0}=u^0,\\
\bu^\prime=\bu-\frac 1c\bv u^0 ,
\eea\label{bellac21}\eeq
which, as we shall see, may be related to the so-called {\it electric}
 limit. The second limit is for spacelike vectors:
\beq\bea{l}
u^{\prime 0}=u^0-\frac 1c\bv\cdot\bu ,\\
\bu^\prime=\bu ,\eea\label{bellac22}
\eeq
and will be associated to the {\it magnetic} limit.
 As it is well known, the space-time coordinates can be described by
 timelike vectors only. Indeed, Eq. (\ref{bellac21}) has the form of
 Galilean inertial space-time transformations:
\beq\bea{l}
\bfx'=\bfx-\bv t,\\
t'=t.\eea\label{galxttrans}\eeq
 Nevertheless, other vectors, such as the four-potential
 and four-current, may transform as one or the other of the two limits. 

An example of the subtlety of non-relativistic 
 kinematical covariance is that it is quite common to neglect to enforce
 the condition that a non-relativistic limit involves not only
 low-velocity phenomena, but also large
 timelike intervals then one obtains different kinematics,
 referred to as Carroll
 kinematics \cite{carroll}. In other terms, a Galilean world is
 one within which units
 of time are naturally much larger than units of space. The existence of events
 physically connected by large spacelike intervals would
 imply loss of causality, among other things. Other such kinematics,
 each one being some limit of the de Sitter kinematics, have
 been classified in \cite{bacry}.

The situation is similar with electric and magnetic fields. One needs to compare
 the module of the electric field $E$ to $c$ times the module of the magnetic
 field, \ie $cB$. When the magnetic field is dominant, Eq.
 (\ref{fieldslorentz}) reduces to a transformation referred to as the
 {\it magnetic limit} of electromagnetism:
\beq\bea{l}
\bE_m'=\bE_m+\bv\times\bB_m,\qquad\qquad E_m<<cB_m,\\
\bB_m'=\bB_m.
\eea\label{fieldsmagnetic}\eeq
The other alternative, where the electric field is dominant, leads to the
 {\it electric limit}:
\beq\bea{l}
\bE_e'=\bE_e,\qquad\qquad E_e>>cB_e,\\
\bB_e'=\bB_e-\frac 1{c^2}\bv\times\bE_e.
\eea\label{fieldselectric}\eeq
Indeed, the approximations $E_e/c>>B_e$ and $v<<c$ together imply that
 $E_e/v>>E_e/c>>B_e$ so that we take $E_e>>vB_e$ in Eq. (\ref{fieldslorentz}).
 Such an analysis of orders of magnitude
  is described in the next section.

From the Galilean transformations of space-time, Eq. (\ref{galxttrans}),
 we find
\beq
\nabla'=\nabla,\qquad\qquad \partial_{t'}=\partial_t+\bv\cdot\nabla.\label{galgradtrans}\eeq
The fields transformations in the {\em magnetic} limit of Eq. (\ref{fieldsmagnetic})
 are clearly compatible with the use of Eq. (\ref{galgradtrans}) together with
 the transformations of the four-potential $(V,\bA)$:
\beq\bea{l}
V_m'=V_m-\bv\cdot\bA_m,\\
\bA_m'=\bA_m,\eea\label{potmag}\eeq
 (note the similarity with Eq. (\ref{bellac22})) where
\beq
\bE_m=-\nabla V_m-\partial_t\bA_m,\qquad\qquad \bB_m=\nabla\times\bA_m.\label{fieldpotmag}\eeq
Similarly, the {\em electric} limit of Eq. (\ref{fieldselectric})
 may be obtained from Eq. (\ref{galgradtrans}) and 
 the transformations of the four-potential:
\beq\bea{l}
V_e'=V_e,\\
\bA_e'=\bA_e-\frac\bv{c^2}V_e.\eea\label{potel}\eeq
This equation is similar to Eq. (\ref{bellac21}).
Now, however, the fields are related to the four-potential by
\beq
\bE_e=-\nabla V_e,\qquad\qquad \bB_e=\nabla\times\bA_e.\label{fieldpotel}\eeq

In parallel with the two possible sets of transformations of the four-potential,
 there are two ways to transform the four-current $(\rho,\bj)$. In the 
{\em magnetic limit}, it transforms like Eq. (\ref{bellac22}):
\beq\bea{l}
\rho_m'=\rho_m-\frac 1{c^2}\bv\cdot\bj_m,\\
\bj_m'=\bj_m,\eea\label{curmag}\eeq
and the continuity equation then reads
\beq
\nabla\cdot\bj_m =0.\label{continuitymagnetic}\eeq

The appearance of an `effective' charge density
 $\rho_m'=\rho_m-\frac 1{c^2}\bv\cdot\bj_m$ is certainly one of the
 salient feature of the magnetic limit. We will refer the interested
 reader to the following works which discussed the effect of this
 effective charge without pointing out its Galilean origin
 for most of them \cite{effective}.

For the {\em electric limit}, it transforms like Eq. (\ref{bellac21}):
\beq\bea{l}
\rho_e'=\rho_e,\\
\bj_e'=\bj_e-\bv\rho_e,\eea\label{curel}\eeq
and the continuity equation has its usual form:
\beq
\nabla\cdot\bj_e+\partial_t\rho_e=0.\label{continuityelectric}\eeq

Finally, Maxwell's equations,
\beq\bea{rcll}
\bfn\ti\bE&=&-\pd_t\bB,\quad & {\rm Faraday},\\ 
\bfn\cdot\bB&=&0,\quad & {\rm Thomson},\\
\bfn\ti\bB&=&\mu_0\bj+\frac 1{c^2} \pd_t\bE,\quad & {\rm Ampere},\\
\bfn\cdot\bE&=&\frac 1{\epsilon_0}{\rho},\quad & {\rm Gauss},
\eea\label{maxwell}\eeq
reduce, in the Galilean limits, to two respective forms.
 As the field transformation
 laws themselves, this fact is not so obvious if one naively takes the limit
 $c\rightarrow\infty$. In the next section, we present an argument based on
 dimensional analysis and orders of magnitude. In Ref. \cite{bellac}, it was
 found that, in the electric limit, the Maxwell equations reduce to:
\beq\bea{l}
\bfn\ti\bE_e=\bze ,\\
\bfn\cdot\bB_e=0 ,\\
\bfn\ti\bB_e-\frac 1{c^2} \pd_t\bE_e=\mu_0\bj_e ,\\
\bfn\cdot\bE_e=\frac 1{\epsilon_0}{\rho_e}.\eea
\label{maxel}\eeq
Clearly, the main difference with the relativistic Maxwell equations is that
 here the electric field has zero curl in Faraday's law.
In the magnetic limit, the Maxwell equations become
\beq\bea{l}
\bfn\ti\bE_m=-\pd_t\bB_m ,\\
\bfn\cdot\bB_m=0,\\
\bfn\ti\bB_m=\mu_0\bj_m ,\\
\bfn\cdot\bE_m=\frac 1{\epsilon_0}\rho_m .\eea\label{maxmag}
\eeq
The displacement current term is absent in Amp\`ere's law.




\section{Recent analyses}


\subsection{Orders of magnitude}

Many errors occur within low-velocity limits of relativistic theories
 when one naively replaces some quantities with zero, rather than carefully 
 comparing various orders of magnitudes involved in the equations. As we
 shall show herafter, we do not require fanciful mathematical tools
 to retrieve the two Galilean limits of electromagnetism of LBLL.
 As discussed by one of us in Ref. \cite{europhys},
 a careful dimensional analysis of the
 fields equations is sufficient for this purpose. Therein, it is argued that the electric
 and magnetic limits may be retrieved by a careful consideration of the order of
 magnitude of the dimensionless parameters:
\beq
\ve\equiv\frac L{cT},\qquad {\rm and}\qquad \xi\equiv\frac j{c\rho},
\label{epsilonxi}\eeq
where $L$, $T$, $j$ and $\rho$ represent the orders of magnitude of
 length, time, current density, and charge density, respectively.
 The Galilean kinematics considered hereafter corresponds to the
 quasistatic limit $\ve<<1$. The other extreme, $\ve>>1$, leads
 to the so-called Carroll kinematics \cite{carroll}. The main undesirable
 feature of this kinematical structure is the loss of causality.

The electric or magnetic character of the Galilean limits
 of electromagnetism is determined by the behavior of the parameter $\xi$.
 From Gauss's law and Amp\`ere's law, Eq. (\ref{maxwell}), we find
 $\frac {cB}E\simeq\frac j{\rho c}$, so that
\[
\frac{cB}E=\xi.
\]
Using this result and Eqs. (\ref{fieldsmagnetic}), (\ref{fieldselectric}), we find:
\beq\bea{l}
\xi>>1:\qquad {\rm magnetic\ limit},\\
\xi<<1:\qquad {\rm electric\ limit}.\eea
\label{xielectricvsmagnetic}\eeq
Returning to Eq. (\ref{epsilonxi}), we see that the magnetic limit correspond to
 the approximation $j>>c\rho$, that is, the spacelike component is larger than 
 the timelike component. This echoes the transformation in Eq. (\ref{bellac22}).
 Conversely, the electric limit correspond to the approximation $c\rho>>j$,
 so that the spacelike component now is much larger than the timelike component.
 This is analogue to Eq. (\ref{bellac21}). 

From Maxwell displacement current term in the Amp\`ere's law, Eq. (\ref{maxwell}),
 we find
\beq
B\simeq\frac{vE}{c^2},
\label{dispcur}\eeq
where $v$ denotes the ratio of orders of magnitude $L/T$. Similarly,
 the magnetic induction term of Faraday's law gives
\beq
E\simeq vB.
\label{faradterm}\eeq
If we substitute this result into Eq. (\ref{dispcur}), we find that the displacement
 current term and the full Faraday's law are compatible only if $v\simeq c$,
 that is, in the Lorentz covariant regime. However, Eq. (\ref{dispcur}) cannot
 be obtained if we drop $\partial_t\bE$ from Amp\`ere's law,
 so that it is compatible with
 the first and third lines of Eq. (\ref{maxel}), i.e. in the electric limit.
 On the other hand, Eq. (\ref{faradterm}) is compatible with
 lines one and three of Eq. (\ref{maxmag}), i.e. the magnetic limit,
 because it does not appear if we
 drop the magnetic induction term $\partial_t\bB$ of Faraday's law
 in line three of Eq. (\ref{maxwell}).

Following the lines of Ref. \cite{Henri90}-\cite{aflb}, we use the
 Riemann-Lorenz formulation of electromagnetism, which relies on
 the potentials as the basic quantities, in order
 to retrieve the two Galilean limits. This is in opposition to
 the Heaviside-Hertz formulation, which is based on the magnetic and
 electric fields \cite{Henri90}-\cite{Alfred}. In terms of potentials,
 the equations of classical electromagnetism read 
\beq\bea{l}
\nabla^2V-\frac 1{c^2}\frac{\partial^2 V}{\partial t^2}=-\frac\rho{\epsilon_0},\qquad
 {\rm Riemann\ equations},\\
\nabla^2\bA-\frac 1{c^2}\frac{\partial^2 \bA}{\partial t^2}=-\mu_0\bj,\eea\label{riemann}
\eeq
\beq
\bfn\cdot\bA+\frac 1{c^2}\frac{\partial V}{\partial t}=0,\qquad {\rm Lorenz\ equation},
\label{lorentz}\eeq
\beq
\frac {\rm d}{{\rm d}t}(m\bv+q\bA)=-q\nabla (V-\bv\cdot\bA),\qquad {\rm Lorentz\ force}.
\eeq

They can be obtained from the Fermi Lagrangian \cite{Fermi}
\begin{equation}
{\cal L}_F = { 1 \over 2} \epsilon_0 c^2 \partial_\mu A_\nu \partial^\mu A^\nu 
\label{lfer}
\end{equation}

We consider the full Lagrangian consisting of a field and a matter part, that is ${\cal L}= {\cal L}_F + {\cal L}_M$.  The Euler-Lagrange equation,
\begin{equation}
\frac{\partial{\cal{L}}}{\partial A_\mu}-\partial_\nu
\left [\frac{\partial{\cal{L}}}{\partial(\partial_\nu A_\mu)}\right ]=0,
\label{eqmotgauge}
\end{equation}
leads to the following equations of motion (using Eq. (\ref{lorentz}) $\partial_\mu A^\mu=0$)
\begin{equation}
\partial_\nu \partial^\nu A^\mu = - \mu_0 j^\mu
\label{wave}
\end{equation}
with
\begin{equation}
- \mu_0 j^\mu = \partial {\cal L}_M / \partial A_\mu .
\label{current}
\end{equation}

The quasistatic approximation, $\ve<<1$, of Eq. (\ref{riemann}) reads
\beq
\nabla^2V\simeq -\frac\rho{\epsilon_0}\qquad\qquad {\rm and}\qquad\qquad
 \nabla^2 {\bf A}\simeq -\mu_0 {\bf j},
\label{potentialvscurrent}\eeq
from which we can define a further dimensionless ratio, 
$\frac {cA}V\simeq\frac j{\rho c}$, so that 
\beq
\frac {cA}V\simeq \xi.
\label{potxi}\eeq
Once again, this echoes our comment following Eq. (\ref{xielectricvsmagnetic}):
in the magnetic limit, the spacelike quantity $cA$ is dominant, whereas in 
 the electric limit, it is the timelike quantity $V$ which dominates.

If we compare the two terms of the Lorenz (not Lorentz \cite{lorenz})
 gauge condition, Eq. (\ref{lorentz}), we find
\beq
\frac{|\nabla\cdot\bA|}{\partial_t V/c^2}\simeq \frac{cT}L\frac{cA}V
\simeq\frac\xi\ve.
\label{xivelorenzcondition}\eeq
In the quasistatic regime, $\ve<<1$, we find therefore two possibilities. If
 $\xi<<1$, like $\ve$, then we are in the electric limit, and the gauge condition
 is similar to the Lorenz condition:
\beq
\bfn\cdot\bA_e+\frac 1{c^2}\partial_tV_e=0.\label{lorenzelectric}
\eeq 
 On the other hand, in the magnetic limit, $\xi>>1$, then we drop $\partial_t V$, so
 that we obtain the Coulomb gauge condition: 
\beq
\bfn\cdot\bA_m=0.
\label{coulomb}
\eeq 

Let us use the orders of magnitude for the four-potential components and obtain thereof
 their Galilean transformations in the magnetic limit, Eq. (\ref{potmag}), and
 the electric limit, Eq. (\ref{potel}). From Eq. (\ref{lorentz4vector}) with
 $u^0=V/c$ and $\bu=\bA$, we find that the scalar potential
 $V$ and the vector potential $\bA$ transform, under a Lorentz transformation, as
\beq\bea{l}
V'=\gamma(V-\bv\cdot\bA),\\
\bA'=\bA-\gamma\frac\bv{c^2}V+(\gamma-1)\frac\bv{\bv^2}\bv\cdot\bA.
\eea\label{potlorentz}\eeq
From the first line of this equation,
 we have $V\simeq vA$, so that we obtain, from Eq. (\ref{potxi}), 
\[
\xi=\frac {cA}V\simeq\frac{cA}{vA}=\frac 1\ve.
\]
Therefore, in the quasistatic limit $\ve<<1$, this equation gives $\xi>>1$,
 so that the first line is compatible with the magnetic limit. Accordingly, this is incompatible with the electric
 limit, for which $\xi<<1$, so that the
 term $\bv\cdot\bA$ must be dropped from the first line of Eq. (\ref{potlorentz})
 in the electric limit. 

A similar argument, applied to the second line of
 Eq. (\ref{potlorentz}), implies $A\simeq\frac{vV}{c^2}$, so that
\[
\xi=\frac {cA}V\simeq\frac{cvV}{c^2V}=\frac vc=\ve.
\]
Unlike the previous case, the quasistatic limit leads to $\xi<<1$, which 
 is compatible with the electric limit only, Eq. (\ref{potel}). This implies
 that, in the magnetic limit, the term $\frac\bv{c^2}V$ must be dropped from the second line of
 Eq. (\ref{potlorentz}), as it is the case in Eq. (\ref{potmag}). 

If we use an entirely similar analysis for the Lorentz transformation of charge and current
 densities, obtained from Eq. (\ref{lorentz4vector}) with $u^0=\rho$ and $\bu=\bj/c$:
\beq\bea{l}
\rho'=\gamma(\rho-\frac 1{c^2}\bv\cdot\bj),\\
\bj'=\bj-\gamma\bv\rho+(\gamma-1)\frac\bv{\bv^2}\bv\cdot\bj,
\eea\label{curlorentz}\eeq
we retrieve the Galilean transformations Eqs. (\ref{curmag}) and (\ref{curel}) for
 the magnetic and electric limits, respectively.

Let us conclude by briefly discussing the continuity equation:
\[
\nabla\cdot\bj+\partial_t\rho=0.
\] 
If we compare the two terms as we have done for the Lorentz gauge condition in
 Eq. (\ref{xivelorenzcondition}), we find
\[
\frac{|\nabla\cdot\bj|}{\partial_t \rho}\simeq \frac{cT}L\frac{j}{c\rho}
\simeq\frac\xi\ve.
\]
If $\xi<<1$, like $\ve$ in the quasistatic regime $\ve<<1$, then we obtain the electric
 limit and we retrieve Eq. (\ref{continuityelectric}). On the other hand, in the magnetic case,
 $\xi>>1$, so that we drop $\partial_t \rho$ and obtain
 thereby Eq. (\ref{continuitymagnetic}).


\subsection{Reduction from $(4,1)$ Minkowski space-time}

Hereafter we review briefly a different approach to
 the Galilean gauge fields \cite{galgauge}.
 It involves a formulation of Galilean invariance based on a reduction from
 a five-dimensional Minkowski manifold to the Newtonian space-time
 \cite{takahashi}-\cite{galileangravity}. 
 The extended space is such that a Galilean boost with
 relative velocity $\bv=(v_1, v_2, v_3)$
 acts on a {\it Galilei-vector} $(\bfx, t, s)$ as
\beq
\begin{array}{l}
\bfx^\prime=\bfx-\bv t,\\
t^\prime=t,\\
s^\prime=s-\bv\cdot\bfx+\frac 12\bv^2t.
\end{array}\label{galileantransf}\eeq
Since $\partial_s$ tranforms like the mass $m$ (see below for a justification), one can see the 
 additional coordinate $s$ as being conjugate to the mass $m$
 since both are invariant under Galilean transformations. 
 $s$ may be seen also as the action per unit mass.
 More about classical and quantum physical interpretations
 of $s$ is in Refs. \cite{galgauge}-\cite{galileangravity}.

The scalar product,
\[
(A|B)=A^\mu B_\mu\equiv {\bf A}\cdot{\bf B}-A_4B_5-A_5B_4,
\]
of two Galilei-vectors $A$ and $B$ is invariant under
 the transformation, Eq. (\ref{galileantransf}).
 This suggests a method to base the tensor calculus on the metric
\beq
g^{\mu\nu}=g_{\mu\nu}=\left ( 
\begin{array}{ccccc}
1 & 0 & 0 & 0 & 0 \\ 
0 & 1 & 0 & 0 & 0 \\ 
0 & 0 & 1 & 0 & 0 \\ 
0 & 0 & 0 & 0 & -1 \\ 
0 & 0 & 0 & -1 & 0
\end{array}
\right ).\label{galileimetric}\eeq
Hereafter we refer to this as the {\it Galilean metric}.

The transformation in Eq. (\ref{galileantransf}) can be
 written in matrix form for the components of any five-vector as
\[
x^{\prime\mu}=\Lambda^{\mu}_{\ \nu} x^\nu,
\]
where $\mu$ denotes the row and $\nu$ the column (so that
 $\Lambda^{\mu}_{\ \nu}$ is the $(\mu\nu)$-entry) or
\[
\left (\bea{c}
x^{'1}\\
x^{'2}\\
x^{'3}\\
x^{'4}\\
x^{'5}\eea\right )=
\left (\bea{ccccc}
1 & 0 & 0 & -v_1 & 0\\
0 & 1 & 0 & -v_2 & 0\\
0 & 0 & 1 & -v_3 & 0\\
0 & 0 & 0 & 1 & 0\\
-v_1 & -v_2 & -v_3 & \frac 12\bv^2 & 1\eea
\right )
\left (\bea{c}
x^{1}\\
x^{2}\\
x^{3}\\
x^{4}\\
x^{5}\eea\right ).
\]
For a five-oneform, this transformation reads
\[
x'_{\mu}=\Lambda_{\mu}^{\ \nu} x_\nu ,
\]
where $\mu$ now denotes the column and $\nu$ the row
 (that is $\Lambda^{\ \nu}_{\mu}$ is the $(\nu\mu)$-entry) , or
\beq
\left (
x'_{1},
x'_{2},
x'_{3},
x'_{4},
x'_{5}\right )=
\left (
x_{1},
x_{2},
x_{3},
x_{4},
x_{5}\right )
\left (\bea{ccccc}
1 & 0 & 0 & v_1 & 0\\
0 & 1 & 0 & v_2 & 0\\
0 & 0 & 1 & v_3 & 0\\
0 & 0 & 0 & 1 & 0\\
v_1 & v_2 & v_3 & \frac 12\bv^2 & 1\eea
\right ).
\label{matrixcontravariant}\eeq

We write the embedding as
\[
(\bx, t)\rightarrow x^\mu=(\bx, t, s),
\]
 as well as the following five-momentum:
\[
p_\mu\equiv -i\pd_\mu=(-i\bfn, -i\pd_t, -i\pd_s),
\]
so that, with the usual identification $E=i\pd_t$, and with 
 $m=i\pd_s$, we obtain
\[\bea{l}
p_\mu=(\bp, -E, -m),\\
p^\mu=g^{\mu\nu}p_\nu=(\bp, m, E).
\eea\]
  Thereupon the mass does not enter as an external
 parameter, but as a remnant of the fifth component of the 
 particle's momentum. Hereafter, the five-momentum operator 
 will act on a massless field, so that
\[
\partial_5A=\partial_sA=0.\]

Now let us set up the five-dimensional quantities that allow us
 to retrieve the two Galilean limits of electromagnetism.
 They are given by defining two embeddings of
 the five-potential:
\[
A_\mu=(\bA, A_4, A_5).
\]
Under the transformation in Eq. (\ref{galileantransf}) 
 its components transform, from Eq. (\ref{matrixcontravariant}), as
\beq\bea{l}
\bA'=\bA+\bv A_5,\\
A_{4'}=A_4+\bv\cdot\bA +\frac 12\bv^2 A_5,\\
A_{5'}=A_5.
\eea\label{atransf}\eeq
Next we write the five-dimensional electromagnetic antisymmetric
 Faraday tensor:
\beq
F_{\mu\nu}\equiv\pd_\mu A_\nu-\pd_\nu A_\mu=\left (
\bea{ccccc}
0 & b_3 & -b_2 & c_1 & d_1 \\
-b_3 & 0 & b_1 & c_2 & d_2 \\
b_2 & -b_1 & 0 & c_3 & d_3 \\
-c_1 & -c_2 & -c_3 & 0 & a \\
-d_1 & -d_2 & -d_3 & -a & 0 \eea
\right ).\label{fcomp}
\eeq
Thus we have
\beq\bea{l}
\bb=\bfn\ti\bA ,\\
\bc=\bfn A_4-\pd_4\bA ,\\
\bd=\bfn A_5-\pd_5\bA ,\\
a=\pd_4A_5-\pd_5A_4.
\eea\label{e0}\eeq

The five-current
\[
j_\mu=(\bj, j_4, j_5),
\]
transforms under the transformation, Eq. (\ref{galileantransf}), as
\beq\bea{l}
\bj'=\bj+\bv j_5,\\
j_{4'}=j_4+\bv\cdot\bj +\frac 12\bv^2 j_5,\\
j_{5'}=j_5.
\eea\label{jtransf}\eeq
The continuity equation takes the form
\beq
\pd^\mu j_\mu=\bfn\cdot\bj-\pd_4 j_5-\pd_5 j_4=0.
\label{continuityequation}\eeq
The five-dimensional Lorenz-like condition takes a similar form:
\beq
\pd^\mu A_\mu=\bfn\cdot\bA-\pd_4 A_5-\pd_5 A_4=0.
\label{lorenzcondition}\eeq

In the presence of sources, the Maxwell equations are
\beq
\pd_\mu F_{\alpha\beta}+\pd_\alpha F_{\beta\mu}+
\pd_\beta F_{\mu\alpha}=0,
\label{maxwell1}\eeq
and
\beq
\pd_\nu F^{\mu\nu}=j^\mu ,
\label{maxwell2}\eeq
so that in terms of the components of $F$ defined in Eq. (\ref{fcomp}),
 we find, from Eq. (\ref{maxwell1}):
\beq\bea{l}
\bfn\cdot\bb=0,\\
\bfn\ti\bc+\pd_4\bb=\bze ,\\
\bfn\ti\bd+\pd_5\bb=\bze ,\\
\bfn a-\pd_4\bd+\pd_5\bc=\bze ,
\eea\label{e1}\eeq
whereas Eq. (\ref{maxwell2}) reduces to
\beq\bea{l}
\bfn\ti\bb-\pd_5\bc-\pd_4\bd=\bj ,\\
\bfn\cdot\bc-\pd_4a=-j_4 ,\\
\bfn\cdot\bd+\pd_5a=-j_5.
\eea\label{e2}\eeq

From $F_{\mu'\nu'}=\Lambda_{\mu'}^\alpha\Lambda_{\nu'}^\beta F_{\alpha\beta}$
 the entries of $F$ in Eq. (\ref{matrixcontravariant}) transform as
\beq\bea{l}
a^\prime=a+\bv\cdot\bd ,\\
\bb^\prime=\bb-\bv\ti\bd ,\\
\bc^\prime=\bc+\bv\ti\bb+\frac 12\bv^2\bd-a\bv-\bv(\bv\cdot\bd) ,\\
\bd^\prime=\bd.
\eea\label{fcomptransf}\eeq

Let us now see how the electric and magnetic limits are contained within the
 previous formulas.


\subsubsection{Electric limit}

 As mentioned previously, the electric limit is characterized
 by four-potential and four-current vectors which are timelike, that is,
 their time component is much larger than the length of their
 spatial components. In the reduction approach, it corresponds
 to defining the embedding of the potentials and currents as
\beq
(\bA_e, V_e)\ho A_e=\left (\bA_e, 0, -\mu_0\epsilon_0V_e\right ) ,
\label{embeddingae}
\eeq
and
\beq
(\bj_e, \rho_e)\ho j_e=\left (\mu_0\bj_e, 0, -\mu_0\rho_e\right ) ,
\label{embeddingje}
\eeq
respectively. 

From Eqs. (\ref{atransf}) and (\ref{embeddingae}) we retrieve Eq. 
 (\ref{potel}). Similarly we obtain Eq. (\ref{curel}) from Eqs. (\ref{jtransf})
 and (\ref{embeddingje}). As for the continuity equation,
 Eq. (\ref{continuityequation}), it becomes Eq. (\ref{continuityelectric}).
 From the first line of Eq. (\ref{e0}), we come to the natural definition:
\[
\bB_e\equiv \bb=\bfn\ti\bA_e.
\]
The electric field is defined as the component $\bd$, so that from
 the third line of Eq. (\ref{e0}) we have
 $\bE_e\equiv \frac 1{\mu_0\epsilon_0}\bd=-\bfn V_e$, as in
 Eq. (\ref{fieldpotel}).
From Eq. (\ref{e0}) we note that 
 $\bc=-\pd_t\bA_e$ and $a=-\frac 1{\mu_0\epsilon_0}\pd_t V_e$.
Then Eq. (\ref{fcomptransf}) leads to Eq. (\ref{fieldselectric}).
 The corresponding Maxwell equations, Eq. (\ref{maxel}),
 are obtained from Eqs. (\ref{e1}) and (\ref{e2}).

Note that the second line of Eq. (\ref{e2}) provides a condition
 similar to Lorenz gauge fixing:
 \[
\bfn\cdot\bA_e=-\mu_0\epsilon_0\pd_t V_e.\]
This expression may be obtained also by substituting
 Eq. (\ref{embeddingae}) into Eq. (\ref{lorenzcondition}).


\subsubsection{Magnetic limit}

This non-relativistic limit is characterized
 by spacelike four-potential and four-current vectors;
 their time component is small compared to the length of their
 spatial components. Hereafter we show that it corresponds
 to defining the embedding of the potentials and currents as
\beq
(\bA_m, V_m)\ho A_m=\left (\bA_m, -V_m, 0\right ),
\label{embeddingam}
\eeq
and
\beq
(\bj_m, \rho_m)\ho j_m=\left (\mu_0\bj_m, -\frac 1{\epsilon_0}\rho_m, 0\right ),
\label{embeddingjm}
\eeq
respectively. 

From Eqs. (\ref{atransf}) and (\ref{embeddingam}) we retrieve Eq. (\ref{potmag}).
 Similarly Eqs. (\ref{jtransf}) and (\ref{embeddingjm}) lead to Eq. (\ref{curmag}),
 and the continuity equation (\ref{continuityequation}) 
 gives Eq. (\ref{continuitymagnetic}), which shows that the current
 $\bj_m$ cannot be related to a convective transport of charge! 
As above, we define the magnetic field as
 $\bB_m\equiv \bb=\bfn\ti\bA_m$, 
and the electric field is now defined as the component $\bc$, so that from
 the second line of Eq. (\ref{e0}) we obtain
 $\bE_m\equiv \bc=-\bfn V_m-\pd_t\bA_m$, as in Eq. (\ref{fieldpotmag}).
 Then Eq. (\ref{fcomptransf}) leads to Eq. (\ref{fieldsmagnetic}).
The Maxwell equations (\ref{maxmag}) are obtained from Eqs. (\ref{e1}) and
 (\ref{e2}). Finally, note that by replacing Eq. (\ref{embeddingam}) into
 Eq. (\ref{lorenzcondition}) we obtain Coulomb's gauge condition:
\[
\bfn\cdot\bA_m=0.
\]



\section{Discussion and examples}

\subsection{Gauge conditions and Galilean electromagnetism}

In this section, we describe the two Galilean limits using
 the Riemann-Lorentz approach, that is, in terms of the scalar
 and vector potentials. The definition ${\bf E}=-\partial_t {\bf A}-\nabla V$
 of Eq. (\ref{fieldpotmag}) takes different forms in the Galilean limit,
 depending on the order of magnitude of each term. This is    
 because the Galilean transformations for the potentials 
 depend on whether we take the electric or magnetic limit.
 Let us evaluate the order of magnitude of the
 ratio between its two terms:
\[
\frac{\partial_t {\bf A}}{\nabla V}\simeq
 \frac{\frac{A}{T}}{\frac{V}{L}}
 \simeq \frac{L}{cT}\frac{cA}{V} \simeq \ve \xi.
\]

In the {\em magnetic} limit, for which $\xi>>1$, this equation
 leads to ${\bf E}_m=-\nabla V_m-\partial_t {\bf A}_m$ as in Eq.
 (\ref{fieldpotmag}). By computing the curl, we find 
 $\partial_t {\bf B}_m=-\nabla \times {\bf E}_m$, as in the
 first line of Eq. (\ref{maxmag}).
 Likewise, in the {\em electric} limit, for which  $\xi<<1$, we can
 neglect $\partial_t\bA$ so that we find 
 ${\bf E}_e\simeq -\nabla V_e$, as stated in Eq.
 (\ref{fieldpotel}). The curl of this expression leads to 
 $\nabla \times {\bf E}_e\simeq {\bf 0}$, as in the first line
 of Eq. (\ref{maxel}).

Let us illustrate how the choice of the gauge condition 
 allows one to retrieve the two sets of ``Galilean
 Maxwell equations'' in terms of the fields,
 as stated by LBLL \cite{bellac}. In the {\em magnetic}
 limit, the condition $\xi>>1$ leads to Eq. (\ref{coulomb}), as
 mentioned earlier. From the definition of $\bB_m$ and using the
 identity
\[
\nabla \times(\nabla \times {\bf A})=\nabla(\nabla
 \cdot{\bf A})-\nabla ^2 {\bf A},
\]
 we obtain 
\[
\nabla \times{\bf B}_m=\nabla\times (\nabla \times {\bf A}_m)=
 \nabla(\overbrace{\nabla\cdot\bA_m}^0)-\nabla^2\bA_m=\mu_0\bj_m,
\]
where the last term follows from Eq. (\ref{potentialvscurrent}).
 This is the third line of Eq. (\ref{maxmag}), where the displacement
 current term is missing. 
 With $\bE_m$ defined as in Eq. (\ref{fieldpotmag}), its divergence
 gives 
\[
\nabla\cdot\bE_m=\nabla\cdot (-\partial_t {\bf A}_m-\nabla V_m)
 =-\partial_t(\overbrace{\nabla\cdot\bA_m}^0)-\nabla^2V_m=
\frac{\rho_m}{\epsilon_0},
\]
where we have utilized again Eq. (\ref{coulomb}). This is the second
 inhomogeneous equation, last line of Eq. (\ref{maxmag}).
  
In the {\em electric} limit, $\xi<<1$ and Eq. (\ref{xivelorenzcondition})
 lead to the Lorenz condition, Eq. (\ref{lorenzelectric}). Proceeding as in
 the magnetic limit, we first calculate the curl of $\bB_e$, which gives us
\[
\nabla \times{\bf B}_e=\nabla\times (\nabla \times {\bf A}_e)=
 \nabla(\overbrace{\nabla\cdot\bA_e}^{-(\partial_t V_e)/c^2})-\nabla^2\bA_e
 =\frac 1{c^2}\partial_t\bE_e+\mu_0\bj_e.
\]
We have utilized Eq. (\ref{fieldpotel}) to define $\bE_e$. This is line
 three of Eq. (\ref{maxel}). Finally, by calculating the divergence
 of $\bE_e$, we find
\[
\nabla\cdot\bE_e=\nabla\cdot (-\nabla V_e)=-\nabla^2 V_e=
\frac{\rho_e}{\epsilon _{0}}, 
\]
where we have used Eqs. (\ref{fieldpotel}) and (\ref{potentialvscurrent}).

Therefore, let us point out forcefully that the choice of
 a gauge condition is dictated by the relativistic versus
 Galilean nature of the problem.
 The Lorenz gauge condition must be chosen in
 the relativistic context as well as in the electric Galilean limit.
 The Coulomb gauge condition can be chosen only in the 
 magnetic limit. For example, this implies that 
 quantization in the Coulomb gauge of light waves is prohibited
  because of Galilean covariance and must be re-examined.
 We refer to a discussion of the physical meaning
 that one can ascribe to the various gauge conditions \cite{aflb}.

From the historical point of view, Galilean electromagnetism has shed
 a new light on the pre-relativity era.
 Indeed, a careful reading of James Clerk Maxwell's
 famous {\it Treatise on Electricity and
 Magnetism} reveals that he has employed the electric limit when dealing with
 dielectrics in his first volume \cite{Maxwell}.
 On the other hand, in his second volume, he used 
 the magnetic limit when dealing with ohmic
 conductors, except toward the end,
 where he introduced into the magnetic limit equations
 the so-called displacement current `by hand'
 in order to demonstrate that light is a transverse electromagnetic
 wave \cite{Maxwell}. But, as we have seen in the particular case of
 the electric limit (and it is also valid in relativity),
 the displacement current follows from choosing the Lorenz gauge whereas
 Maxwell (wrongfully!) kept the Coulomb gauge
 within the relativistic context for the fields.
 This difficulty enticed Hertz and Heaviside to abandon 
 potentials and to cast Maxwell's equations in terms of 
 fields rather than potentials. Albert Einstein read
 Hertz' papers on the topic and subsequently
 employed Maxwell equations in terms of
 the fields (the Heaviside-Hertz formulation) whereas Henri
 Poincar\'e employed Maxwell equations in terms of the potentials
 (the Riemann-Lorenz formulation) by adopting the
 Lorenz condition in a relativistic context \cite{darrigol1}.

\subsection{Magnetoquasistatics (MHD) and electroquasitatics (EHD)}

Experts working on magnetohydrodynamics (MHD) and experts of 
 electrohydrodynamics (EHD) might be 
 surprised to realize that they actually work with
 different sets of approximations of the Maxwell
 equations where retardation (and, therefore,
 waves) has been neglected.
 Indeed, the displacement current
 is negligible in MHD (as in Eq. (\ref{maxmag})),
 whereas in EHD the electric field
 has a vanishing curl (Eq. (\ref{maxel}).
 Hence, effects that are important in MHD becomes
 marginal in EHD, and vice versa. 

Melcher has greatly clarified
 these facts by disjoining the electroquasistationary
 approximation used in EHD
 and the magnetoquasistationary approximation used in MHD \cite{melcher}
 Section 3.2 of Melcher and Haus \cite{melcher} shows that
 the underlying equations are precisely the same as the Galilean
 limits, Eqs. (\ref{maxel}) and (\ref{maxmag}).
 His main argument relies on the comparison
 between three characteristics time scales ($\sigma$ is the electric conductivity):
 (1) the magnetic diffusion time $\tau _m =\mu _0 \sigma L^2$,
 (2) the charge relaxation time $\tau _e =\epsilon _0 /\sigma$ and
 (3) the wave transit time $\tau _{em} =L/c=\sqrt{\tau _e \times \tau _m}$, which
 is the square root of the product of the two former time scales.
 By definition, the transit time is always between
 the magnetic and electric times.
 For example, in the magnetic limit,
 the charge relaxation time scale is very small and
 the magnetic field does have enough time
 to diffuse inside the Ohmic carrier. It is straightforward
 to see that the electroquasistatic
 of Melcher corresponds to the electric limit
 whereas the magnetic limit is just the magnetoquasistatic.
 Hence, a large amount of our technology is based on
 Galilean electromagnetism as
 soon as waves are neglected.
 
\subsection{The Faraday tensor and its dual}

It is well known that in special relativity, the Faraday
 tensor, defined in Eq. (\ref{fcomp}):
\[
F_{\mu\nu}\equiv \partial_\mu A_\nu-\partial_\nu A_\mu,
\] 
(in this section, $\mu,\nu=0,1,2,3$)
 and its dual:
\[
^*F_{\mu\nu}=\frac 12\epsilon^{\mu\nu\rho\sigma}F_{\rho\sigma},
\]
 do have the same physical meaning. This is not the case
 within Galilean electromagnetism. Indeed, as pointed out by
 Earman and subsequently discussed by Rynasiewicz, the Galilean
 tranformations of the Faraday tensor and its dual lead to the
 electric or the magnetic limit, respectively \cite{Earman}.
 The effect of this duality operation amounts to exchanging
 $\bE$ and $\bB$ as follows:
 \[
{\bf{E}}\rightarrow c{\bf{B}}\quad {\rm and}\quad
 {\bf{B}}\rightarrow -{\bf{E}}/c.
\]
 One recovers very easily the magnetic and electric limits,
 Eqs. (\ref{fieldsmagnetic}) and (\ref{fieldselectric}),
 by applying the duality transformations
 directly to the electric transformations of the field
 in order to get the magnetic transformations, and vice versa.

It is also noted in Ref. \cite{Earman} that the field transformations of
 the magnetic limit are obtained when $\bE$ and $\bB$ are expressed
 in terms of `covariant', or
 $\left (\begin{array}{cc} 0\\ 2\end{array}\right )$, tensor $F_{\mu\nu}$,
 whereas the electric limit is obtained when the fields transformations
 are calculated by using the `contravariant', or
 $\left (\begin{array}{cc} 2\\ 0\end{array}\right )$, tensor $F^{\mu\nu}$.
 Let us illustrate it briefly, with
\[
A^\mu=\left (\frac Vc,\bA\right ),\qquad A_\mu=\left (\frac Vc,-\bA\right ),
\]
as well as
\[
\partial^\mu=\left (\frac 1c\partial_t,-\nabla\right ),\qquad
 \partial_\mu=\left (\frac 1c\partial_t,\nabla\right ).
\]

The magnetic limit rests on the relation
\[
F'_{\mu\nu} = \Lambda_\mu^{\ \rho} \Lambda_\nu^{\ \sigma}\; F_{\rho\sigma},
\]
where the Galilean transformation matrix $\Lambda_\mu^{\ \nu}$
 is defined by the four-gradient transformation,
 $\partial'_\mu=\Lambda_\mu^{\ \nu}\partial_\nu$,
 so that
\[
\Lambda_\mu^{\ \nu}=\left (
\begin{array}{cccc}
1 & \frac{v_x}c & \frac{v_y}c & \frac{v_z}c \\
0 & 1 & 0 & 0 \\
0 & 0 & 1 & 0 \\
0 & 0 & 0 & 1 \end{array}\right ). 
\]
The index $\mu$ denotes the line of each entry.
We find, for example,
\[
\begin{array}{rcl}
 \frac{E'_x}c=F'_{01}&=& \Lambda_0^{\ \mu} \Lambda_1^{\ \nu}\; F_{\mu\nu},\\
 &=& \Lambda_0^{\ 0}F_{01} + \Lambda_0^{\ 2}F_{21} +
 \Lambda_0^{\ 3}F_{31},\\
 &=& \frac 1c (E_x+v_yB_z-v_zB_y),
\end{array}
\]
and
\[
-B'_z=F'_{12}=\Lambda_1^{\ \mu}\Lambda_2^{\ \nu}F_{\mu\nu}=-B_z,
\] 
which is Eq. (\ref{fieldsmagnetic}). 

The electric limit transformations follows from 
\[
F'^{\mu\nu} = \Lambda^\mu_{\ \rho} \Lambda^\nu_{\ \sigma}\; F^{\rho\sigma}.
\]
The transformation matrix $\Lambda^\mu_{\ \nu}$
 is now defined by the coordinate transformation,
 $x^\mu=\Lambda^\mu_{\ \nu}\; x^\nu$, with
 $x^\mu=(ct,x,y,z)$, so that
\[
\Lambda^\mu_{\ \nu}=\left (
\begin{array}{cccc}
1 & 0 & 0 & 0 \\
-\frac{v_x}c & 1 & 0 & 0 \\
-\frac{v_y}c  & 0 & 1 & 0 \\
-\frac{v_z}c  & 0 & 0 & 1 \end{array}\right ). 
\]
Again, the first index $\mu$ denotes the matrix line. 
For instance, we compute
\[
-\frac{E'_x}c=F'^{01}=\Lambda^0_{\ 0}\Lambda^1_{\ \nu}F^{0\nu}=-\frac{E_x}c,
\] 
and
\[
\begin{array}{rcl}
 {B'_z}=-F'^{12}&=&-\Lambda^1_{\ \mu} \Lambda^2_{\ \nu}\; F^{\mu\nu},\\
 &=& -\Lambda^1_{\ 0}F^{02} - \Lambda^2_{\ 0}F^{10} - F^{12},\\
 &=& B_z-\frac 1{c^2}(v_xE_y-v_yE_x),
\end{array}
\]
which is Eq. (\ref{fieldselectric}).

\subsection{Quantum mechanics}

In 1990, Dyson published a demonstration of Maxwell equations due
 to Richard Feynman \cite{dyson}. This demonstration, which
 relied on Lagrangians and quantum mechanics,
 dated back to the forties and had
 remained hitherto unpublished. However, the proof
 was believed to be incomplete because
 Feynman had discussed only the homogeneous Maxwell equations,
 given by the first two lines in Eq. (\ref{maxwell}):
\[
\bfn\ti\bE=-\pd_t\bB,\qquad \bfn\cdot\bB=0.
\]
 In 1999, Brown and
 Holland revisited this demonstration and they have noted the
 Schr\"odinger equation admitted external potentials only if  
 they were compatible with the {\it magnetic} limit of LBLL 
 and, therefore, with the Coulomb gauge condition \cite{brownholland}.
 It is clear from Eq. (\ref{maxmag}) that the homogeneous Maxwell
 equations given above are valid only within the magnetic limit,
 because the electric field has zero curl in the electric limit, Eq.
 (\ref{maxel}). This is a consequence of the Galilean magnetic limit of the
 four-potential which does enters into Schr\"odinger equation.
 Let us recall the statement of Brown and Holland more precisely.
 The Schr\"odinger equation with external fields $V(\bx,t)$ and
 $\bA(\bx,t)$:
\[
\ri\hbar\partial_t\Psi(\bx,t)=
\frac 1{2m}(-\ri\hbar\nabla-\bA(\bx,t))^2\Psi(\bx,t)+V(\bx,t)\Psi(\bx,t),
\]  
is covariant under Galilean transformation, Eq. (\ref{galxttrans}), with
 $\Psi(\bx,t)\rightarrow\Psi'(\bx',t')$, $V(\bx,t)\rightarrow V'(\bx',t')$,
 and $\bA(\bx,t)\rightarrow \bA'(\bx',t')$, if
\[
\begin{array}{l}
{\Psi'(\bx',t')}={\rm const}\; \exp[(\ri/\hbar)
 (-m\bv\cdot\bx+\frac 12m\bv^2t+\phi(\bx,t))]\; \Psi(\bx,t),\\
V'(\bx',t')=V(\bx,t)-\partial_t\phi(\bx,t)-\bv\cdot
 (\bA(\bx,t)+\nabla\phi(\bx,t)),\\
\bA'(\bx',t')=\bA(\bx,t)+\nabla\phi(\bx,t),\end{array}
\]
where $\phi(\bx,t)$ is some scalar function. Note that the factor
 in front of $\Psi(\bx,t)$ transforms like the parameter $s$ in
 Eq. (\ref{galileantransf}). If we choose $\phi(\bx,t)=0$, then
 the equations above reduce to Eq. (\ref{potmag}), which represents
 the Galilean transformations of the potentials in the magnetic limit. 
 
 This point agrees with later results by
 Vaidya and Farina \cite{vaidyafarina}.  In a subsequent study,
 Holland and Brown have shown that Maxwell equations admit an electric limit only if
 the source is a Dirac current \cite{brownholland2}. In addition,
 they have proved that the Dirac equation admits
 both Galilean limits, just like Maxwell equations,
 corroborating thereby earlier
 results by L\'evy-Leblond \cite{levyleblond1967}. What
 Feynmann did not (actually, could not) realize is that he had derived only
 the part of Maxwell equations compatible with both
 Galilean relativity and quantum mechanics, that is, the
 magnetic limit and, hence, the homogeneous equations.

\subsection{Superconductivity}

Superconductivity also enters into the realm of the magnetic limit, because
 it selects the Coulomb gauge condition as a necessary consequence of
 Galilean covariance. Indeed, the well-known London equation states that the current
 density is proportional to the vector potential (the star denotes Cooper pairs): ${\bf p }=m^* {\bf v}+q^* {\bf A}={\bf 0}$. As a matter of fact, there is a perfect transfert of electromagnetic momentum to kinetic momentum. Hence, contrary to what is usually stated, gauge invariance is not broken by superconductivity since the Coulomb gauge condition is implied. Moreover, the Meissner effect
 can be explained by starting with Amp\`ere's equation written as
 $\nabla\times\bB=\mu_0 \bj$, that is, without displacement current term as
 in the magnetic case, third line of Eq. (\ref{maxmag}). Hence, this expression (or more directly $\nabla^2 {\bf A}\simeq -\mu_0 {\bf j}$ in the Riemann-Lorenz formulation) together with $\nabla\cdot\bA=0$ and London equation, leads to solutions (in one dimension $x$) of the type $A\simeq\exp{-{\rm const}\; x}$ so that the vector potential (hence the magnetic field) only penetrates the superconductor to a depth $1/{\rm const}$.

 Superconductivity cannot be associated to a symmetry breaking of gauge invariance but is magnetic Galilean covariant. This unusual statement has been recently advocated by Martin Greiter by a different approach \cite{greiter}. As a matter of fact, it is the global U(1) phase
 rotation symmetry that is spontaneously violated. A striking consequence 
 is that the Higgs mechanism for providing mass to particles becomes doubtful,
 since it was believed to be analogous to the assumed symmetry
 breaking of gauge invariance in superconductivity.

\subsection{Electrodynamics of continuous media at low velocities}

One often finds in textbooks that a dielectric in motion is characterized
 by the presence of a motional polarization given by the following formula \cite{melba}:
\beq
\bP'=\epsilon_0 \chi (\bE+\bv\ti\bB).
\label{polarization}\eeq
where $\chi$ is the dielectric susceptibility. For example,
 H.A. Lorentz utilized it in order to derive the Fresnel-Fizeau
 formula at first order (see p. 174-176 of Ref. \cite{melba} and Ref.
 \cite{adrien}). We will see that this formula can be misleading.
 
 Moreover, in 1904, Lorentz claimed that a moving magnet could become electrically
 polarized \cite{Lorentz}. In 1908, Einstein and Laub noted that
 Minkowski transformations for the fields and the excitations \cite{Minkowski} predict
 that a moving magnetic dipole develops an electric dipole
 moment \cite{EL}. It would be interesting to reexamine these
 predictions in the light of Galilean electromagnetism within
 continuous media. Indeed, if one starts
 from the Minkowski transformations relating the polarization and the
 magnetization \cite{Minkowski}, one would expect two Galilean limits:
 one with ${\bf{M}}' = {\bf{M}}$ and ${\bf{P}}' = {\bf{P}}-{\bf{v}}\times {\bf{M}}/c^2$
 and the other with ${\bf{M}}' = {\bf{M}}+{\bf{v}} \times {\bf{P}}$
 and ${\bf{P}}' = {\bf{P}}$ (see Chapter 9 of Ref. \cite{melba}).
 
 We can derive from the Riemann-Lorenz formulation Maxwell's equations within continuous media following O'Rahilly \cite{Alfred}. More directly, one can infer the form of Maxwell equations
 in continuous media by mimicking the vacuum case by either supressing
 the displacement current or the Faraday term.

Maxwell equations in continuous media are covariant under the
 Poincar\'{e}-Lorentz transformations \cite{Minkowski}:
\[\bea{l}
\bfn\cdot\bB=0,\\
\partial_t\bB=-\bfn\ti\bE,\\
\bfn\cdot\bD=\rho,\\
\bfn\ti\bH=\bj+\partial_t\bD.\eea
\]
In continuous media, the constitutive laws relate the excitation $\bD$,
 the field $\bE$ and the polarization $\bM$:
\[\bea{l}
\bD=\epsilon_0\bE+\bP,\\
\bB=\mu_0\bH+\bM.\eea
\]
These relations are valid in both Galilean and
 Einsteinian relativity \cite{Minkowski}. Let us now turn to
 the electromagnetic laws when we take into account the
 motion of a medium at low velocity \cite{melcher}.
 
 First, we recall the Galilean transformations for the differentiation operators:
 $$
\nabla ' \times ( \cdots ) = \nabla  \times ( \cdots )
$$
$$
\nabla '\cdot( \cdots ) = \nabla \cdot( \cdots )
$$
$$
\nabla '( \cdots ) = \nabla ( \cdots )
$$
$$
{{\partial {\bf{a}}'} \over {\partial t'}} = {{\partial {\bf{a}}'}
 \over {\partial t}} + ({\bf{v}}\cdot\nabla ){\bf{a}}'
$$
$$
\nabla  \times ({\bf{a}} \times {\bf{b}}) = ({\bf{b}}\cdot\nabla ){\bf{a}} -
 ({\bf{a}}\cdot\nabla ){\bf{b}} + {\bf{a}}(\nabla\cdot{\bf{b}}) -
 {\bf{b}}(\nabla\cdot{\bf{a}})
$$

If ${\bf{a}} = {\bf{v}}$ and ${\bf{b}} = {\bf{A}}'$, then:
$$
{{\partial {\bf{A}}'} \over {\partial t'}} = {{\partial {\bf{A}}'} \over {\partial t}}
 + {\bf{v}}(\nabla \cdot{\bf{A}}') - \nabla  \times ({\bf{v}} \times {\bf{A}}')
$$
Let us apply these transformations to the two Galilean limits of Maxwell
 equations expressed for a continuous medium. We write first Maxwell equations
 in a frame of reference R' moving at a relative velocity $\bv$
 with respect to another frame R:\\
\begin{table}[htbp]
\begin{center}
\begin{tabular}{ccc}
\underline{Magnetic Limit} & \hspace{1in} & \underline{Electric Limit}\\
$\nabla ' \times {\bf{H}}' =  {\bf{j}}'$ & & $\nabla ' \times {\bf{E}}' = {\bf{0}}$\\
$\nabla '\cdot{\bf{B}}' = 0$ & & $\nabla '\cdot{\bf{D}}' = \rho '$\\
$\nabla '\cdot{\bf{j}}' = 0$ & & $\nabla '\cdot{\bf{j}}' +
 {{\partial \rho '} \over {\partial {\rm{t'}}}} = 0$\\
$\partial _{t'} {\bf{B}}' = {\rm{ - }}\nabla ' \times {\bf{E}}'$ & & $\nabla ' \times {\bf{H}}'
 = {\bf{j}}' + {{\partial {\bf{D}}'} \over {\partial t'}}$\\
\end{tabular}
\end{center}
\label{default}
\end{table}

We apply the spatial and temporal Galilean derivatives, so that in R', we find:
\begin{table}[htbp]
\begin{center}
\begin{tabular}{ccc}
\underline{Magnetic Limit} & \hspace{1cm} & \underline{Electric Limit}\\
$\nabla  \times {\bf{H}}' = {\bf{j}}'$ & & $\nabla  \times {\bf{E}}' = {\bf{0}}$\\
$\nabla \cdot{\bf{B}}' = 0$ & & $\nabla \cdot{\bf{D}}' = \rho '$\\
$\nabla \cdot{\bf{j}}' = 0$ & & $\nabla \cdot({\bf{j}}' + \rho '{\bf{v}}) + {{\partial \rho '} \over
 {\partial {\rm{t}}}} = 0$\\
$\partial _t {\bf{B}}' = {\rm{ - }}\nabla  \times ({\bf{E}}' - {\bf{v}} \times {\bf{B}}')$
 & & $\nabla  \times ({\bf{H}}' + {\bf{v}} \times {\bf{D}}') = {\bf{j}}' + \rho {\bf{v}} +
 {{\partial {\bf{D}}'} \over {\partial t}}$
\end{tabular}
\end{center}
\label{default}
\end{table}

Hence, we deduce the fields transformations:
\begin{table}[h!]
\begin{center}
\begin{tabular}{ccc}
\underline{Magnetic Limit} & \hspace{1in} & \underline{Electric Limit}\\
${\bf{B}} = {\bf{B}}'$ & & ${\bf{E}} = {\bf{E}}'$\\
 & & $\rho  = \rho '$\\
${\bf{j}} = {\bf{j}}'$ & & ${\bf{j}} = {\bf{j}}' + \rho '{\bf{v}}$\\
${\bf{H}} = {\bf{H}}'$ & & ${\bf{H}} = {\bf{H}}' + {\bf{v}} \times {\bf{D}}'$\\
${\bf{E}} = {\bf{E}}' - {\bf{v}} \times {\bf{B}}'$ & & ${\bf{D}} = {\bf{D}}'$\\
${\bf{M}} = {\bf{M}}'$ &  & ${\bf{P}} = {\bf{P}}'$\\
${\bf{P}} = {\bf{P}}'+{\bf{v}}\times {\bf{M}}'/c^2$ & & ${\bf{M}} = {\bf{M}}'-{\bf{v}} \times {\bf{P}}'$
\end{tabular}
\end{center}
\label{default}
\end{table}
\newpage
As suspected, the effects predicted by Lorentz and Einstein \& Laub in continuous media are of galilean origin and not relativistic !

In addition, we can derive easily the boundary conditions for moving media with ${\bf{n}} =
 {\bf{n}}_{1 \to 2}$ the unit vector between the two media denoted by 1 and 2
\begin{table}[h!]
\begin{center}
\begin{tabular}{ccc}
\underline{Magnetic Limit} & \hspace{1cm} & \underline{Electric Limit}\\
${\bf{n}} \times ({\bf{H}}_2  - {\bf{H}}_1 ) = {\bf{K}}$ & & ${\bf{n}} \times ({\bf{E}}_2  - {\bf{E}}_1 ) = 0$\\
${\bf{n}}\cdot({\bf{B}}_2  - {\bf{B}}_1 ) = {\bf{0}}$ & & ${\bf{n}}\cdot({\bf{D}}_2  - {\bf{D}}_1 ) = \sigma $\\
${\bf{n}}\cdot({\bf{j}}_2  - {\bf{j}}_1 ) + \nabla _\Sigma  \cdot{\bf{K}} = {\bf{0}}$ & &
 ${\bf{n}}\cdot({\bf{j}}_2  - {\bf{j}}_1 ) + \nabla _\Sigma
  \cdot{\bf{K}} = v_n (\rho _2  - \rho _1 ) - \partial _t \sigma $\\
${\bf{n}} \times ({\bf{E}}_2  - {\bf{E}}_1 ) = v_n ({\bf{B}}_2  - {\bf{B}}_1 )$ & & 
 ${\bf{n}} \times ({\bf{H}}_2  - {\bf{H}}_1 ) = {\bf{K}} + v_n {\bf{n}}
 \times [{\bf{n}} \times ({\bf{D}}_2  - {\bf{D}}_1 )]$
\end{tabular}
\end{center}
\label{default}
\end{table}

$\bf{K}$ is the surface current, $\sigma$ the surface charge, $\Sigma$ the
 surface separating both media, and $v_n$ the projection of the relative
 velocity on the normal of $\Sigma$.

The formula, Eq. (\ref{polarization}), used by Lorentz is
 not compatible with Galilean relativity. However,
 the electric field and the magnetic field which create the polarization in Fizeau
 experiment come from a light wave so that
 Lorentz was right to use this formula after all,
 even though there is no contradiction with the electric limit
 formula ${\bf{P}}' = {\bf{P}}=\epsilon _0 \chi {\bf E'}=\epsilon _0 \chi {\bf E}$.
  
\subsection{Electrodynamics of moving bodies at low velocities}

Galilean electromagnetism raises severe doubts concerning our current understanding of
 the electrodynamics of moving media. Indeed, several experiments,
 like the ones by Roentgen \cite{Roentgen},
 Eichenwald \cite{Eichenwald}, Wilson \cite{Wilson}, Wilson and
 Wilson \cite{WW}, Trouton and Noble \cite{trouton}, etc., were believed
 to corroborate special relativity. As we will demonstrate for the Trouton-Noble
 experiment, there is not always a need
 for special relativity because the typical relative
 velocity in these experiments is well below the speed of light.
 Then, the question is whether the above mentioned
 experiments be explained by either the electric limit, the magnetic
 limit or a combination of both.

\subsubsection{The Trouton-Noble experiment}

The Trouton-Noble's experiment is thought to be the electromagnetic analogue of the optical
 Michelson and Morley experiment \cite{trouton}. It was designed in order to show whether one
 can observe a mechanical velocity of the ether if one considers that the
 luminiferous medium should be a medium whose parts can be followed mechanically.
 Like the null result of Michelson-Morley's optical experiment,
 the Trouton-Noble experiment was negative in the sense
 that one was not able to detect either an absolute motion with respect to the ether,
 or a partial entrainment, as believed by other various theories.

In 1905, Albert Einstein suggested to consider the ether as superfluous
 since the experiments were not able to detect a mechanical motion of it. Others, like
 H. Poincar\'{e} and H.A. Lorentz, were reluctant to abandon the notion of ether as the
 bearer of the electromagnetic field, despite the fact that they have adopted the
 relativity principle. However, in 1920, Einstein recoursed to the
 ether as the bearer of the metric allowing the propagation of gravitational waves,
 at a conference in Leyden. Today, even though the ether is a banished
 word in modern science, one can use it as did the older Einstein in order to
 describe the vacuum with physical (though not mechanical) properties.

Before the advent of special relativity, Hertz, Wien, Abraham, Lorentz, Cohn and others have
 used the transformations given in Eq. (\ref{wrongfieldtrans}), which is an incoherent
 mixture of the electric and magnetic Galilean limits \cite{darrigol2}. As mentioned
 previously, these expressions do not even obey the group property of composition of transformations.

Trouton and Noble expected a positive effect when a charged capacitor is in motion
 with an angle $\theta$  between the plates and the velocity \cite{trouton}.
 Indeed, the electric field in
 the frame of the capacitor generates a magnetic field in the ether frame
 ($\bv$ is the absolute velocity):
\[
\bB'=-\frac 1{c^2}\bv\ti\bE
\]
that is,
\[
B'=\frac 1{c^2}vE\sin\theta.
\]
Hence, there is a localization of magnetic energy density inside a volume d${\cal V}$:
\[
{\rm d}W=\frac 12\frac{B'}{\mu_0}{\rm d}{\cal V}= \frac 12\frac{v^2}{c^2}\epsilon_0E^2
\sin^2\theta\ {\rm d}{\cal V}.
\]
With the volume of the capacitor given by $Sl$, the total energy between the plates is
\[
W=\frac 12\frac{v^2}{c^2}\epsilon_0E^2\sin^2\theta\ Sl.
\]
If one denotes as $V=E/l$ the difference of potential between the plates, then the
 capacitor is submitted to the electrical torque
\[
\Gamma=-\frac{{\rm d}W}{{\rm d}\theta}=-\frac{\epsilon_0}2\frac{V^2S}{l}
\frac{v^2}{c^2}\sin(2\theta),
\]
which is maximal for $\theta=45^\circ$ and zero for $\theta=90^\circ$. Hence, the
 plates must be perpendicular to the velocity. One does {\em not} observe such an effect 
 in practice. 

In order to understand what is wrong with the above demonstration, we first recall
 that the electric limit transforms as in Eq. (\ref{fieldselectric}). A straightforward
 application of these transformations is to note that the Biot-Savart law follows  
 from the Coulomb law associated to the electric transformation of the magnetic field.
 Contrary to the transformations used by Lorentz and Trouton-Noble, they do respect
 the group additivity. Besides, these transformations are only compatible with the
 approximate set of Maxwell equations where the time derivative in the Faraday 
 equation vanishes, as in Eq. (\ref{maxel}). Hence, we can derive easily the approximate
 Poynting's theorem within the electric limit from this set:
\beq
\partial_t\left (\frac 12\epsilon_0E^2\right )+
\bfn\cdot\left (\frac{\bE\ti\bB}{\mu_0}\right )\simeq -\bj\cdot\bE.
\label{poyntingel}\eeq
As we can see, the energy density is of electrical origin only. Hence, no electric energy
 associated with the motional magnetic field can be taken into account within the
 electric limit since it is of order $v^2 /c^2$ with respect to the static, or
 quasistatic, electric one. Thus, the Trouton-Noble experiment does not show any effect
 as soon as we are in the realm of the electric limit. We recall that the electric limit 
 is such that the relative velocity is small compared to the velocity of light $c$, and the 
 order of magnitude of the electric field is large compared to the product of $c$ by
 the magnetic field. Of course, for larger velocities, special relativity is needed and
 we must take into account additional mechanical torque due to lenght contraction as
 usual now in order to have a negative result (no torque).
 
The conclusion of Trouton and Noble is rather illuminating
 concerning the fact that they did consider the energy of the motional
 magnetic field to be the source of the negative result \cite{trouton}:\\

{\em ``There is no doubt that the result is a purely negative one. As the
 energy of the magnetic field, if it exists (and from our present point
 of view we must suppose it does), must come from somewhere, we are
 driven to the conclusion that the electrostatic energy of a capacitor
 must dimininish by the amount $1/2 \epsilon _0 E^2 v^2/c^2$ , when
 moving with a velocity $v$ at right angles to its electrostatic lines
 of force where $1/2 \epsilon _0 E^2$ is the electrostatic energy.''}\\

Conversely, a solenoid/magnet in motion will not create a motionnal magnetic torque because the
 magnetic energy associated with the motional electric field is negligible compared
 to the magnetic energy of the static, or quasistatic, magnetic field.
 
 \subsubsection{``Einstein asymmetry''}
 
In his famous article on the electrodynamics of moving media, Albert Einstein pointed out 
 the importance of whether or not one should ascribe energy to the fields when
 dealing with motion \cite{Einstein}. We reproduced here the introduction of his paper:\\
 
{\em ``It is known that Maxwell's electrodynamics -as usually understood at the
present time- when applied to moving bodies, leads to asymmetries which do
not appear to be inherent in the phenomena. Take, for example, the reciprocal
electrodynamic action of a magnet and a conductor. The observable phenomenon
here depends only on the relative motion of the conductor and the
magnet, whereas the customary view draws a sharp distinction between the two
cases in which either one or the other of these bodies is in motion. For if the
magnet is in motion and the conductor at rest, there arises in the neighbourhood
of the magnet an electric field with a certain definite energy, producing
a current at the places where parts of the conductor are situated. But if the
magnet is stationary and the conductor in motion, then no electric field arises in the
neighbourhood of the magnet. In the conductor, however, we find an electromotive
force to which in itself there is no corresponding energy, but which gives
rise -assuming equality of relative motion in the two cases discussed- to electric
currents of the same path and intensity as those produced by the electric
forces in the former case.''}\\

Like our discussion of the Trouton-Noble experiment, the magnetic Poynting's
 theorem can explain why one cannot ascribe an energy to the motional electric field
 in Einstein's thought experiment:
\[
\partial_t\left (\frac{B^2}{2\mu _0}\right )+
\bfn\cdot\left (\frac{\bE\ti\bB}{\mu_0}\right )\simeq -\bj\cdot\bE.
\]
This is the magnetic analogue of Eq. (\ref{poyntingel}).

By applying the Lorentz transformation that he had derived in the
 kinematical analysis of his article
 to the `Maxwell' equations (in fact, he used the Heaviside-Hertz formulation,
 unlike Poincar\'{e}, who used the Riemann-Lorenz formulation in his
 relativity article),
 Einstein replaced Lorentz's explanation: \\

{\em ``1. If a unit electric point charge is in motion in an electromagnetic field,
there acts upon it, in addition to the electric force, an Òelectromotive forceÓ
which, if we neglect the terms multiplied by the second and higher powers of
v/c, is equal to the vector product of the velocity of the charge and the magnetic
force, divided by the velocity of light. (Old manner of expression.)''}\\

\noindent{by the now famous special relativity explanation (valid for all velocities):}\\

{\em ``2. If a unit electric point charge is in motion in an electromagnetic field,
the force acting upon it is equal to the electric force which is present at the
locality of the charge, and which we ascertain by transformation of the field to
a system of co-ordinates at rest relatively to the electrical charge. (New manner
of expression.)''}\\

Then he concluded:\\

{\em ``The analogy holds with Òmagnetomotive forces.Ó We see that electromotive
force plays in the developed theory merely the part of an auxiliary concept,
which owes its introduction to the circumstance that electric and magnetic forces
do not exist independently of the state of motion of the system of co-ordinates.
Furthermore it is clear that the asymmetry mentioned in the introduction
as arising when we consider the currents produced by the relative motion of a
magnet and a conductor, now disappears. Moreover, questions as to the ÒseatÓ
of electrodynamic electromotive forces (unipolar machines) now have no point.''}\\

However, we point out forcefully that the Galilean magnetic transformations of the
 electromagnetic field are sufficient to explain the magnet and conductor thought
 experiment of Albert Einstein as Lorentz covariance is not necessary \cite{norton}.
 It means that the second postulate (constancy of the velocity of light)
 used by Einstein was not necessary to explain the thought experiment. Only the relativity
 postulate and the magnetic Galilean transformations are necessary as the usual relative
 velocity of a real experiment is much more inferior to the velocity 
 of light. Hence for low velocities regime, we proposed the following removal of Einstein's asymmetry:\\
 
{\em ``3. If a unit electric point charge is in motion in an electromagnetic field,
the force acting upon it is equal to the electric force which is present at the
locality of the charge, and which we ascertain by a Galilean magnetic transformation of the field to
a system of co-ordinates at rest relatively to the electrical charge. (New manner
of expression only valid for low velocities.)''}\\

Einstein was right to replace Lorentz explanation as Lorentz thought that the cross
 product of the velocity and the magnetic field was not an electric field that's
 why the latter called it in particular the electromotive field. But Einstein did
 not see that the same cross product was an effective electric field due to a
 magnetic Galilean transformations. Wolfgang Pauli provided a resolution of the asymmetry
 in his textbook on electrodynamics but he only assumed that his
 calculations were a first order approximation of the relativistic
 demonstration without acknowledging the existence of a Galilean approximation
 as the magnetic limit \cite{Pauli}. Hence, Einstein's procedure to remove the
 asymmetry is completely valid despite the fact that special relativity is not
 necessary to remove it but only sufficient. Hence, ironically, the thought experiment
 who leaded Albert Einstein to special relativity could have been explained by Galilean
 relativity only with the use of the magnetic limit...\\
 
 As pointed out a long time ago by Keswani and Kilminster \cite{Keswani}, Maxwell
 did resolve Einstein's asymmetry within the formalism of the magnetic limit:\\
 
{\em  ``In all phenomena relating to closed circuits and the current in them, it is
 indifferent whether the axes to which we refer the system be at rest or in motion''.}\\

Indeed,\\

{\em ``The electromotive intensity is expressed by a formula of the same type, wheter
 the motions of the conductors be referred to fixed axes or to axes moving in
 space, the only differences between the formulae being that in the case of
 moving axes the electric potential $V$ must be changed into $V'=V-{\bf v}.{\bf A}$.
 In all cases in which a current is produced in a conducting circuit, the
 electromotive force is the line-integral $e=\int _C {\bf E'}.d{\bf s}$ 
 taken round the curve. The value of V disappears from this integral, so that
 the introduction of $-{\bf v}.{\bf A}$  has no influence on its value.''}

\section*{Concluding remarks}

One century after the relativity revolution and more than thirty
 years after the forgotten work of L\'evy-Leblond and Le Bellac,
 Galilean electromagnetism is becoming a field of actual research
 as we can explain much more simply scores of experiments involving
 the electrodynamics of moving media without the sophisticated formalism
 of special relativity.

 Moreover, for slow velocities, it is now obvious that special relativity's effects such as the length
 variation cannot explain (as it was believed so far) these
 experiments since it is negligible. In the realm of mechanics,
 what would have happened if Newton was born after Einstein? We
 are exactly in this situation with respect to electromagnetism.

\section*{Acknowledgement}

\noindent 
M.deM. is grateful to NSERC (Canada) for financial support. G.R. thanks
 E. Guyon, B. Jech, A. Domps, M. Le Bellac, O. Darrigol, R. Kofman, J. Rubin, 
 J. Reignier and Y. Pierseaux for fruitful discussions on electromagnetism
 and relativity. G.R. was financially supported by a CNRS postdoctoral
 grant (S.P.M. section 02, France) during his stay in Nice.



\begin{thebibliography}{99}
\bibitem{riemannlorenz} In this paper, we emphasize the 
 Riemann-Lorenz approach to electromagnetism. Therein the
 central role is played by the vector and scalar potentials, unlike the
 Heaviside-Hertz approach, which rather relies on the
 fields themselves. For a justification, see Ref. \cite{Alfred} below.
\end{thebibliography}
\end{document}